# Spatial resolution measurement of Triple-GEM detector and diffraction imaging test at synchrotron radiation

Y.L. Zhang,[a,b,c] H.R. Qi,[b,c] Z.W. Wen,[a,b,c] H.Y. Wang,[b,c,d] Q. Ouyang,[b,c] Y.B. Chen,[b,c] J. Zhang[b,c] and B.T. Hu[a*]

[a] *Lanzhou University,*
   *730000, Lanzhou, China*
[b] *State Key Laboratory of Particle Detection and Electronics,*
   *100049, Beijing, China*
[c] *Institute of High Energy Physics, Chinese Academy of Sciences,*
   *100049, Beijing, China*
[d] *Graduate University of Chinese Academy of Sciences,*
   *100049, Beijing, China*
   *E-mail:* qihr@ihep.ac.cn

ABSTRACT: A triple-GEM detector with two-dimensional readout is developed. The detector provides high position resolution for powder diffraction experiments at synchrotron radiation. Spatial resolution of the detector is measured in the lab using a $^{55}$Fe X-ray source. A resolution of about 110 μm FWHM is achieved. The energy resolution is better than 27% for 5.9 keV X-rays. The detector's validity under illumination of photons in particular energy range is verified using a Cu X-ray tube. Imaging of the head of a wire stripper with X-ray tube demonstrates its imaging ability. A diffraction imaging experiment using the sample of powder $SiO_2$ is successfully carried out at 1W2B laboratory of Beijing Synchrotron Radiation Facility (BSRF). Different diffraction rings are clearly seen under various X-ray energies.

KEYWORDS: Gas Electron Multiplier (GEM); spatial resolution; synchrotron radiation; powder X-ray diffraction.

---

[*] Corresponding author.

## Contents

## 1. Introduction

Synchrotron radiation is the electromagnetic radiation emitted when electrons or other charged particles are forced by magnetic fields to follow curved trajectories [1]. Thanks to its peculiar characteristics, such as broad and continuous spectral range, high intensity, high brightness, natural narrow angular collimation and pulsed time structure, synchrotron radiation is one of the most powerful tools for investigating the properties of matter in many different fields [2]. From X-ray tubes to storage rings of different generations, the average brightness have enhanced by 4–13 orders of magnitude. The existing X-ray detectors, however, either do not offer the time and spatial resolutions needed or are not capable of dealing with the tremendous photon rates produced by synchrotron radiation facilities, and are often the bottleneck in the experiment [3]. The unique detector for every application does not exist and the detector has to be optimized for a particular application.

Gas-filled detectors have long been used as position sensitive instruments since the first synchrotron experiments [4]. Two dimensional multi-wire proportional chambers (MWPCs) [5], an outstanding representative of gaseous detector, were equipped on many synchrotron beamlines about 40 years ago [6]. But were progressively replaced by the image plate and then the scintillator/CCD [7] and now the so-called pixel detectors. This is mainly because MWPCs cannot match the counting rate and position resolution requirements. Gaseous detectors, however, have their unique advantages and are irreplaceable under specific conditions. Wide-angle and small angle X-ray scattering (WAXS, SAXS) are fields where gas-filled detectors are almost indispensable [4]. The newly developed micro pattern structure gaseous detectors with a collective name of Micro Pattern Gaseous Detectors (MPGDs), make it possible to develop new generation of gaseous detectors with distinguished spatial resolution, high rate capability and large sensitive area [8-10]. Investigations of MPGDs used in synchrotron radiation as an alternative of MWPCs have been carried out during the past decades [11-14]. Introduced in 1997 in high energy physics [15], the gaseous electron multiplication (GEM) detector is widely applied in nuclear and particle experiments around the world [16-20]. Cascaded GEM detectors can provide high spatial resolution together with high rate capability. Its usage in synchrotron experiment for time-resolved structure research has also been successfully studied [21].

With the purpose of narrowing the gap between the Beijing Synchrotron Radiation Facility (BSFR) and its detectors, we constructed a two-dimensional single-photon-counting gaseous detector based on triple-GEM for BSRF. In this paper, the detector construction and tests in the lab with radioactive source and X-ray tube are presented. After that an imaging test on synchrotron radiation beam is carried out. The results are described which display a positive attitude toward applying two-dimensional GEM detector to X-ray diffraction experiments with X-rays in the energy range 8–18keV.

## 2. Experimental setup

Figure 1 shows a schematic cross section of the detector. The two-dimensional detector is made up of an entrance window (above the drift cathode, not shown in the figure), a drift electrode, three GEM foils from CERN, and a readout printed circuit board with strips on x- and y-axis.

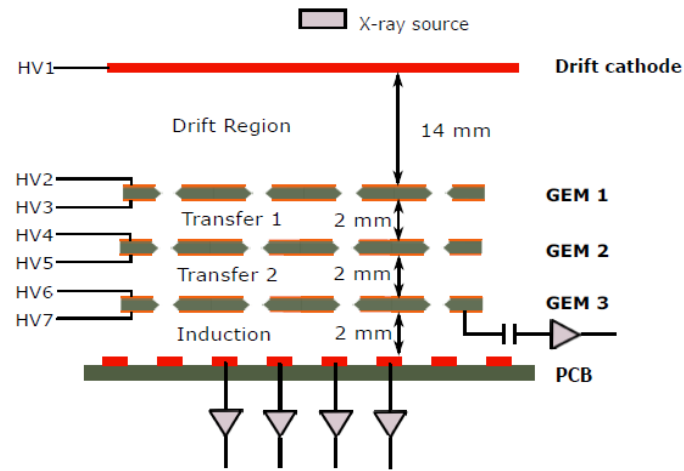

Figure 1. Schematic cross section of the triple-GEM detector set-up.
The sensitive area has a size of 200 mm ×200 mm.

The device has three identical GEM electrodes and an anode separated by 2 mm gaps, the drift gap is 14 mm thick. An individual powering scheme was applied with CAEN N471A high voltage units to bias high voltages on each electrode. The electric field of the transfer regions between the individual GEMs was set to 3 kV/cm and that between the undermost GEM and the readout structure to 4 kV/cm. The voltage across the GEM foils was ranged from 330--340V. A homogeneous drift field between the topmost GEM and the drift cathode was set to 1.2 kV/cm. The working gas was a mixture of $Ar/CO_2(90/10)$ at room temperature and atmospheric pressure.

The X-ray photons enter the detector through the entrance window and the subjacent drift cathode. Inside the drift region, one single photon produces a few hundred primary electrons in the gas volume via photoelectric effect. Under the effect of drift field, electrons drift towards the GEM. The GEM foil is a 50 μm thick Kapton with 5 μm thick copper cladding on both sides and pierced with a high density of holes. When appropriate voltage difference is applied on both sides, a high electric field is generated inside the holes. Electrons drift into the GEM holes are therefore get amplified. Since the electrons exit the back of the topmost GEM, a second and a third GEM placed after the first one provide additional stages of amplification. After multiplication the charge cluster hits the readout strips. The induced charges are collected by the readout strips then get processed by the following electronics. Since ions produced during the avalanche process are collected by the copper electrodes of GEM foils, the induced signals on the readout strips are therefore totally come from collection of electrons, without a contribution from the slow positive ions, making the device potentially very fast.

Figure 2 shows the design of the two-dimensional readout strips. The readout strips in X direction are 193 μm wide at 752 μm intervals. There are pads with a size of 356 μm×356 μm connected with each other in Y direction and their strip pitch is 457 μm. The difference in strip widths is to improve signal sharing between x and y strips, to ensure a homogeneous charge distribution between neighboring strips. The total number of strips in X and Y directions are 267 and 437 respectively. Each strip is connected with one electronic channel to process the signal.

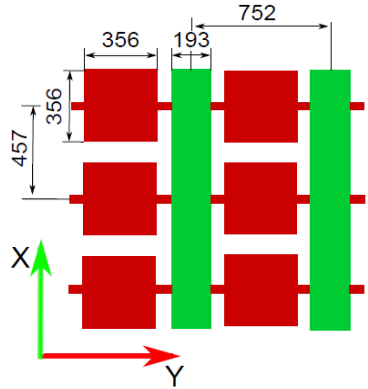

Figure 2. Two-dimensional readout strips design of triple-GEM detector, unit in μm.

Input signals get pre-amplified by a charge amplifier and shaped by a CR-(RC)$^3$ filter with 810 ns peaking time. The detector is aimed at a dedicated imaging device, so this peaking time is a compromise between high rate capability and spatial resolution. With this relatively long peaking time, the induced charge can be fully collected to enhance signal-to-noise ratio. The amplifier outputs are sampled at a rate of 40 MHz. When a trigger signal arrives, samples of the signal amplitude on each channel are read out and digitized by a 10-bit ADC. The amplitude of the output signal, proportional to the initial number of electron-ion pairs produced by the incident photon, and time information are extracted. Signals coming from the bottom side of the undermost GEM foil are used to generate a trigger for the data-acquisition system.

## 3. Measurement of spatial resolution and energy with $^{55}$Fe

The intrinsic spatial resolution (resolution will be used for short later in this paper) is of great importance for an imaging detector [22]. Different criteria lead to varied definitions of resolution but they are equivalent in practice. Full width at half maximum (FWHM) and the standard deviation $\sigma$ are two criteria often used to describe the resolution of imaging detectors. In fact, when the consideration function is of a normal distribution, $FWHM = 2\sqrt{2ln2}\sigma$. During the test, X-ray beams were collimated by a slit with a width of 20 μm before entering the detector volume. A $^{55}$Fe X-ray source with a characteristic energy of 5.9 keV was used to measure the spatial resolution of the triple-GEM detector. Photons emitted into the detector volume through the slit and the induced signals are obtained on related readout strips. The X(Y) coordinate of the hit was determined by an algorithm calculating the center-of-gravity of the charge. Method to calculate the spatial resolution of GEM detector can be found in the reference [23].

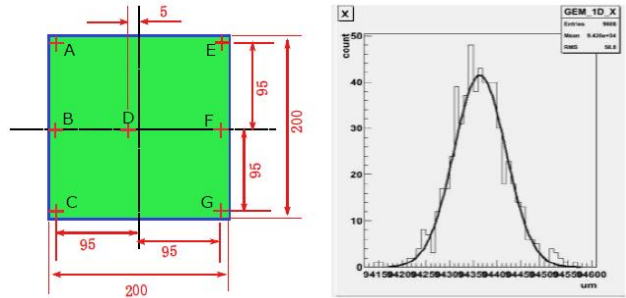

Figure 3. Left: Position of the selected points to be measured; Right: Position resolution with 20 μm slit@Ar/$CO_2$(90/10) of point D.

In the experiment, seven points are selected within the active area of GEM detector to measure the spatial resolution; their positions are shown in figure 3. As an example, this figure also shows the spatial resolution measurement result in X direction of point D. The spatial resolution test result of the seven points in both X and Y directions are summarised in table 1, where σ means the standard deviation of Gaussian fit as shown in figure 3.

Table 1. Measured spatial resolution of different points

| Point | Y (μm) | | X (μm) | |
| :---: | :---: | :---: | :---: | :---: |
| | σ | FWHM | σ | FWHM |
| A | 47.0 | 110.7 | 85.0 | 200.2 |
| B | 49.0 | 115.4 | 80.0 | 188.4 |
| C | 85.1 | 200.4 | 87.2 | 205.3 |
| D | 49.5 | 116.6 | 82.1 | 193.3 |
| E | 65.6 | 154.5 | 98.0 | 230.7 |
| F | 60.0 | 141.3 | 93.0 | 219.0 |
| G | 54.0 | 127.1 | 68.0 | 160.1 |

As the table shows, there are discrepancies of spatial resolution between each point because they are different in location. On the other hand, the relative positions of the slit and readout strips varied from point to point during the test. This gave rise to variant charge distribution on the readout strips, and then discrepancies of spatial resolution between each point. Moreover, for each point, the resolution result in Y direction is superior to the results in X direction. This is because the strip pitch in Y direction is smaller than the strip interval in X axis as shown in figure 3.

The energy calibration of the detector and electronics was carried out using the $^{55}$Fe source ($E_X$ = 5.9 keV). In the operation of argon based gas mixture, a typically pulse height spectrum of GEM detector contains one major peak corresponding to the 5.9 keV X-rays and an escape peak at lower pulse heights corresponds to the ionization energy of an electron from argon's K-shell, as shown in figure 4. The x-axis represents the collected signal charge, which relies on the calibration of the preamplifier-amplifier-ADC chain by injection of different known charges and measuring the corresponding ADC values. The spectrum is fitted with a Gaussian distribution to find the mean and the sigma for the two peaks. The energy resolution is measured to be 25.6% (FWHM). The measured charge can be further transformed into the X-ray energy according to the known energies of escape peak and full energy peak of $^{55}$Fe which is the case of the next test with X-ray tube.

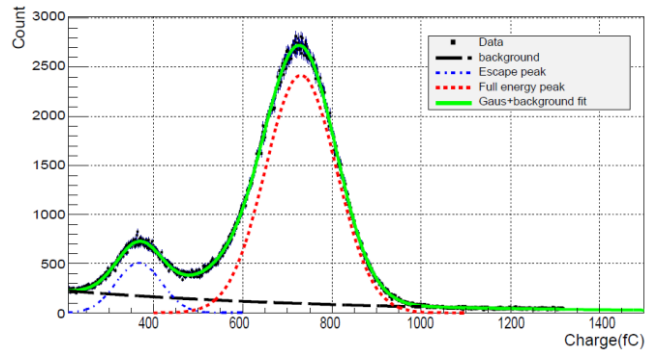

Figure 4. Pulse height spectrum on $^{55}$Fe. The spectrum is fitted with a Gaussian distribution to find the mean and the sigma for the two peaks. The ADC number is converted into collected charge. With the known photon energy from $^{55}$Fe source, it can be further converted to photon energy.

## 4. Energy range test with X-Ray tube

The detector is aimed at X-ray imaging in the energy range 8-18 keV. So after the operation of the detector had been verified by analysing its response to a $^{55}$Fe source, it was tested with an X-ray tube in the lab to verify its validity under illumination of photons in this particular energy range. The experiment used an X-ray tube with Cu target for illumination. X-rays are generated in a cathode ray tube by heating a filament to produce electrons, accelerating the electrons toward a Cu target by applying a voltage, and bombarding the Cu material with electrons. When the X-ray tube is operated at a more high voltage, more energetic photons will be emitted. The maximum energy of the X-rays is related to the voltage at which the X-ray tube is operated. Figure 5 shows the continuous energy spectrum of X-rays emitted by the X-ray tube, operated at 8 kV. When the X-ray tube was operated at 20 kV, the energy spectrum of emitted X-rays is shown in figure 6.

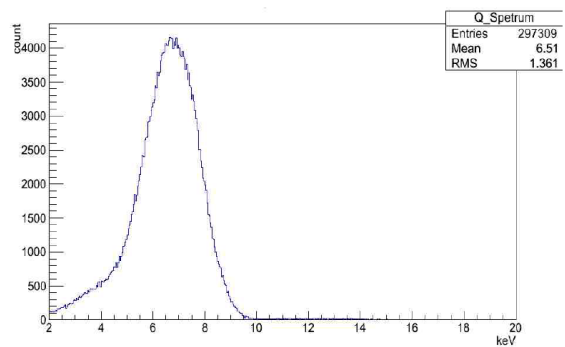

Figure 5. Energy spectrum of the X-rays emitted by an X-ray tube with a Cu target, operated at 8 kV.

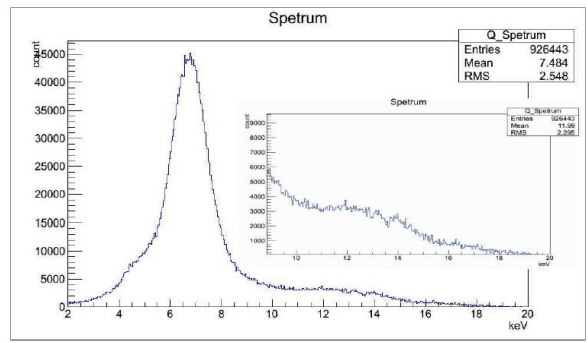

Figure 6. Energy spectrum of the X-rays emitted by an X-ray tube
with a Cu target, operated at 20 kV. Insert: enlarged scale in X-ray
energy region 10-20 keV.

The smooth, continuous curve is due to bremsstrahlung, and the spike is characteristic K lines ($K_\alpha$ and $K_\beta$ are indistinguishable because of the energy resolution of the detector) for Cu atoms. Spectrum of X-rays in the range 10-20 keV is also shown in detail with enlarged scale in figure 6. It can be concluded that the GEM detector has an energy response capability in an energy range from 8 keV to 18 keV. Figure 7 shows a X-ray absorption image of the head of a wire stripper with X-ray tube operated at 20 kV.

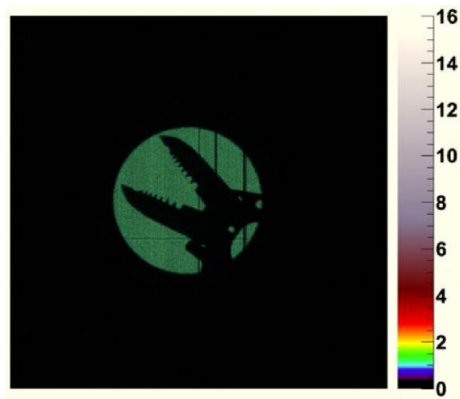

Figure 7. Imaging of the head of a wire stripper with X-ray tube
operated at 20 kV.

## 5. Diffraction imaging with synchrotron radiation X-ray test beam

The imaging experiment is carried out in the synchrotron beamline (BSRF, 1W2B). The X-ray energy of the 1W2B station is adjustable, with a range from 8 keV to 18 keV and the spot size is (H×V) 1×0.6 mm$^2$. At 1W2B laboratory of synchrotron radiation source, the experimental setup for testing is shown in figure 8.

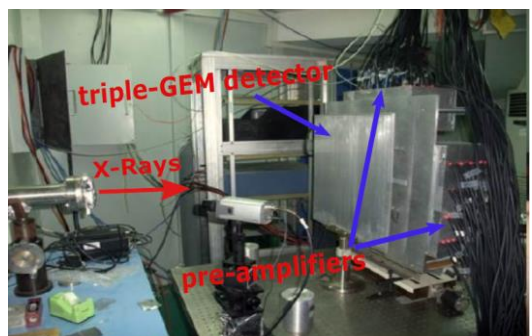

Figure 8. Photograph of view of the GEM detector and the electronics.

The beam piping, slit, SiO$_2$ sample and detector were arranged at the same center height. Taking into consideration of protection of the detector, a lead block is arranged.

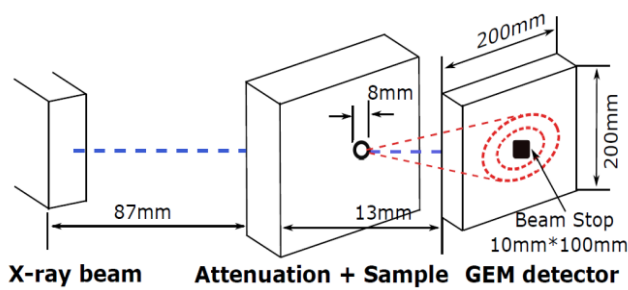

Figure 9. A schematic diagram of diffraction experimental setup on synchrotron radiation X-ray beam.

Figure 9 shows a schematic diagram of diffraction experimental setup. Parallel synchrotron X-rays with wavelength of λ from the beam pipe irradiated the powder SiO$_2$ polycrystalline sample and produced diffraction at certain angles [24] according to Bragg's law:
$$2d\sin\theta = \lambda,$$
where d is the interplanar distance between lattice planes in the crystalline sample and θ is half of the scattering angle.

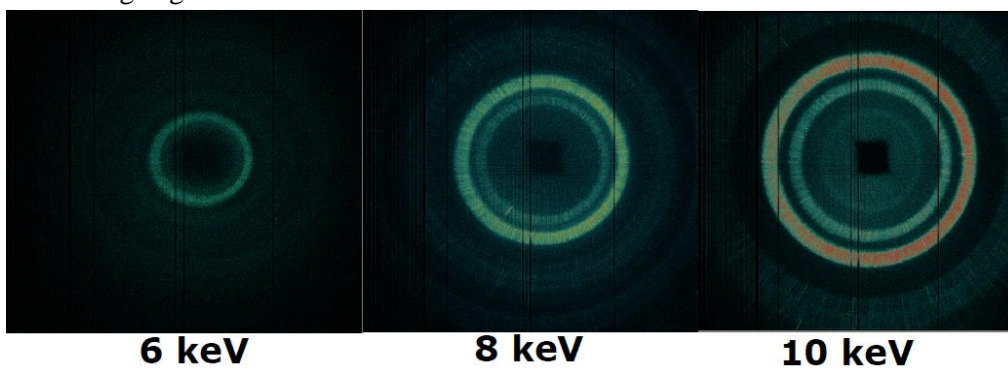

Figure 10. Diffraction imaging of powder SiO$_2$ polycrystalline irradiated by X-ray beams with the energy of 6 keV, 8 keV and 10 keV.

The tested diffraction imaging is shown in figure 10. Black square area in the center indicates the lead block. The sample was irradiated with X-ray beams for three different energies. According to Bragg's law, the interplanar spacing d determines 2θ with a given wavelength. Powder sample contains tens of thousands of randomly oriented crystallites. As X-rays are scatters, there will be a cone corresponds to a single Bragg angle 2θ. With the triple-GEM area detector, a diffraction ring thus get recorded which indicating a specific interplanar spacing d. As is shown in the figure, different rings correspond to various interplanar spacings of the powder sample. As can be seen from figure 10, each diffraction ring has a width which is the contribution of detector caused parallax error aberration [25, 26]. This error can be corrected with appropriate method as discussed in [27] and we will do it in our follow-on work. In spite of this, our experiment demonstrated the effectiveness of the triple-GEM detector in diffraction imaging with synchrotron radiation X-ray test beam.

## 6. Conclusions

In this paper, we have developed a triple-GEM detector for imaging at synchrotron radiation. It is a two-dimensional readout area detector. Tests of this detector have been carried out with a $^{55}$Fe X-ray source and X-ray tube in Ar-$CO_2$ (90-10) gas mixture. The energy resolution of this detector is measured to be 25.6% (FWHM). Spatial resolution has been measured at seven different positions of the detector. The resolution is measured to be 111-231 μm in FWHM depending on the position of the selected points. Tests with an X-ray tube in the lab are carried out to verify the detector's validity under illumination of photons in particular energy range. An X-ray absorption image is presented. Finally a diffraction imaging test is implemented at synchrotron radiation beam and the results are satisfactory.

## Acknowledgments

We gratefully acknowledge the contributions of all members at 1W2B laboratory of Beijing Synchrotron Radiation Source and we appreciate Y.J. WANG for preparation of the samples.
This work was supported by the National Natural Science Foundation of China (Grant No. 11675197) and National Key Programme for S&T Research and Development (Grant No. 2016YFA0400400).